\documentclass[twocolumn,showpacs,amsmath,amssymb]{revtex4}
\usepackage{graphicx}
\usepackage{bm}

\begin{document}

\title{Quantum transfer matrix method for one-dimensional disordered electronic systems}

\author{ L. P. Yang$^1$, Y. J. Wang$^2$, W. H. Xu$^1$, M. P. Qin$^3$, and T. Xiang$^{3,1}$}

\address{
$^1$Institute of Theoretical Physics, Chinese Academy of
Sciences, P.O. Box 2735, Beijing 100080, China\\
$^2$Department of Physics, Beijing Normal University, Beijing
100875, China\\
$^3$Institute of Physics, Chinese Academy of Sciences, P.O. Box 603,
Beijing 100080, China }

\date{\today}

\begin{abstract}

We develop a novel quantum transfer matrix method to study
thermodynamic properties of one-dimensional (1D) disordered
electronic systems. It is shown that the partition function can be
expressed as a product of $2\times2$ local transfer matrices. We
demonstrate this method by applying it to the 1D disordered Anderson
model. Thermodynamic quantities of this model are calculated and
discussed.

\end{abstract}


 \maketitle

\section{Introduction}

For real solids, the perfect periodicity is an idealization, while
the imperfections are of great importance for transport properties.
The broken translational symmetry makes the system deviate from the
extended Bloch waves behavior, and in some cases, to localized
states. As pointed out in Ref.~\cite{Lee}, we can not adopt the
model of ordered systems to understand disordered materials. The
concept of Anderson localization~\cite{Anderson58} and the
correlation effect~\cite{Altshuler} among electrons in a disordered
medium are two important ingredients in the understanding of
disordered systems.

As a minimal Hamiltonian for independent electrons in a disordered
potential, the Anderson disordered model remains difficult to be
understood in finite temperature. The disorder itself invalidates
conventional analytical methods. No proper perturbation parameter
can be chosen to deal with disordered Hamiltonian although
perturbation theory was applied to calculate the conductivity in
weak disorder limit. Considerable efforts focus on Anderson
localization and corresponding metal-insulator transition. In
particular, much insight was gained from the scaling
analysis\cite{Wegner,Abrahams}.

In contrast to higher dimensional cases, 1D models are often
accessible to obtain detailed theoretical (analytical and numerical)
results. However, in the disordered case, it is intractable to make
calculation in the thermodynamic limit by analytical methods even in
1D. The aim of this work is to develop a novel method to resolve
this technical problem. Our method avoids direct diagonalization of
the Hamiltonian and allows the thermodynamic limit to be explored
directly and accurately. The key point lies in the fact that we can
exploit the full translational symmetry in the Trotter (imaginary
time or inverse temperature) direction after trading the evolution
in real space direction with the Trotter one.

It should be noted that the transfer matrix introduced in the
present scheme is not the one usually used in the study of
disordered systems~\cite{Brandes}. Our starting point is to express
the partition function of the system analytically in terms of the
transfer matrix, rather than to use it to trace the eigenvalues or
wave functions.

We will take the disordered Anderson model as an example to
demonstrate how the quantum transfer matrix method works. The
Hamiltonian is defined by
\begin{equation}
H=-\sum_{i}t_i(c_i^\dag c_{i+1}+ {\rm h.c.}) +
\sum_i(U_i-\mu)c_i^\dag c_{i}, \label{eq:H}
\end{equation}
where $t_i$ is the hopping integral between two adjacent sites,
$c_i(c_i^{\dagger})$ is a fermion annihilation(creation) operator at
site $i$, $U_i$ is the diagonal disordered potential, and $\mu$ is
the chemical potential. $t_i$ and $U_i$ can take random values
satisfying some distributions, respectively.

\section{Quantum Transfer matrix method}

As in the quantum transfer matrix renormalization
group(TMRG)~\cite{Tao,Tao1,Xiang} method, we first separates $H$
into two parts, $H=H_1+H_2$, with each part a sum of commuting
terms:
\begin{equation}
H_1=\sum_{i={\rm odd}}h_{i,i+1}, ~~~ H_2=\sum_{i={\rm
even}}h_{i,i+1},
\end{equation}
where \begin{equation} h_{i,i+1}= -t_i(c_i^\dag c_{i+1}+ {\rm h.c.})
+ (U_i-\mu)n_i~.
\end{equation}

The TMRG uses the second-order approximation of the Trotter-Suzuki
formula\cite{Trotter,Suzuki}
\begin{equation}
Z=\mathrm{Tr}\,(e^{-\beta H})= \mathrm{Tr}\,(V_1V_2)^M + O
(\epsilon^2), \label{eq:Z}
\end{equation}
where $\beta = 1/ k_BT$ and $T$ is the temperature. $\beta$ is then
divided into $M$ parts uniformly, $\varepsilon=\beta/M$ and $M$ is
the Trotter number.
\begin{eqnarray}
V_1=e^{-\varepsilon H_1}=\prod\limits_{i={\rm odd}}v_{i,i+1}, \nonumber \\
V_2=e^{-\varepsilon H_2}=\prod\limits_{i={\rm even}}v_{i,i+1},
\end{eqnarray}
where $v_{i,i+1}$ are the local evolution operators defined by
$v_{i,i+1}=e^{-\varepsilon h_{i,i+1}}$. By inserting $2M$ identities
\begin{equation}
\sum|n_1\cdot\cdot\cdot n_N\rangle\langle n_1\cdot\cdot\cdot n_N|=1
\end{equation}
between the neighboring $V_1$ and $V_2$ operators in (\ref{eq:Z})
and labeling successively the complete bases with $l\in[1,2M]$ (so
called imaginary time's slices), the partition function can then be
expressed as
\begin{eqnarray}
 Z&=&\lim_{\varepsilon\rightarrow0}\sum_{\{n_i^l\}}\prod^{M}_{l=1}\langle n_1^{2l-1}~\cdot\cdot\cdot~
n_N^{2l-1}|V_1|n_1^{2l}~\cdot\cdot\cdot~ n_N^{2l}\rangle\nonumber \\
&&~~~~~~~~~~~~~~~~\langle n_1^{2l}~\cdot\cdot\cdot~
n_N^{2l}|V_2|n_1^{2l+1}~\cdot\cdot\cdot~ n_N^{2l+1}\rangle\nonumber \\
&=&\lim_{\varepsilon\rightarrow0}\sum_{\{n_i^l\}}\prod^{M}_{l=1}(v_{1,2}^{2l-1,2l}~\cdot\cdot\cdot~
v_{N-1,N}^{2l-1,2l})\nonumber
\\&&~~~~~~~~~~~~~~~~(v_{2,3}^{2l,2l+1}~\cdot\cdot\cdot~v_{N,1}^{2l,2l+1})
,\label{eq:oriZ}
\end{eqnarray}
where $v_{i,i+1}^{l,l+1}=\langle
n_i^l,n_{i+1}^l|v_{i,i+1}|n_i^{l+1},n_{i+1}^{l+1}\rangle$ represents
the matrix element of $v_{i,i+1}$. The subscripts $i$ and
superscripts $l$ for $n$ and $v$ stand for the coordinates in the
real and Trotter directions, respectively. If we collect all
$v_{i,i+1}^{l,l+1}$ with the same $(i,i+1)$, the partition function
can be re-expressed as the column quantum transfer
operators~\cite{Suzuki,Betsuyaku}:
\begin{equation}
Z=\lim_{\varepsilon\rightarrow0}\mathrm{Tr}\{T_{1,2}T_{2,3}\cdot\cdot\cdot
T_{N,1}\}.\label{eq:part}
\end{equation}
In Eq.~(\ref{eq:part}), there exist $N$ site-dependent column
transfer operators $T_{i,i+1}$, which are defined by a product of
$M$ local transfer operators,
\begin{eqnarray}
T_{2i-1,2i}&=&\prod_{l}\tau_{2i-1,2i}^{2l-1,2l}\nonumber\\
T_{2i,2i+1}&=&\prod_{l}\tau_{2i,2i+1}^{2l,2l+1}, \label{eq:TaTb}
\end{eqnarray}
where the matrix element of the local transfer operator $\tau$ is
defined by
\begin{equation}
\tau_{i,i+1}^{l,l+1}\equiv\langle
n_i^l,1-n_i^{l+1}|v_{i,i+1}|1-n_{i+1}^{l},n_{i+1}^{l+1}\rangle.\label{eq:TH}
\end{equation}

\begin{figure}[ht]
\includegraphics [width=0.85\columnwidth, bb=14 403 552
789] {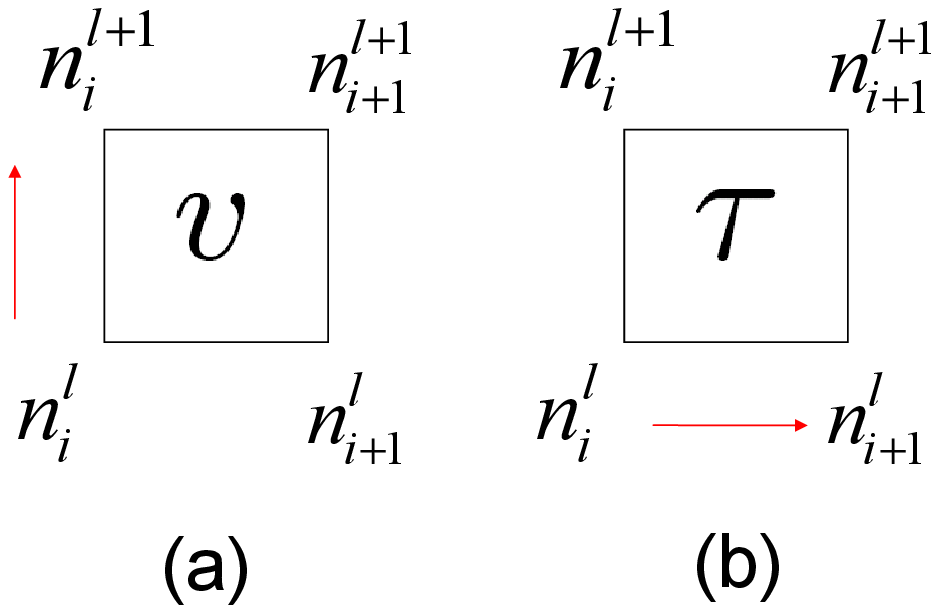} \caption{(Color online) Graphical representation
of the local evolution operator $v$ (a) and the local transfer
matrix (b) along the Trotter and real space directions,
respectively.} \label{fig:transfer}
\end{figure}
The evolution of this matrix is illustrated in
Fig.~(\ref{fig:transfer}). The reason for labeling the basis state
by $1-n^l_i$ is to ensures the conservation of the total occupation
number of two adjacent sites in the Trotter direction. Since
$h_{i,i+1}$ conserves the total occupation number at sites $i$ and
$i+1$, subspaces $ \left\{ |1,1\rangle \right\} $, $\left\{
|0,0\rangle \right\} $ and $\left\{ |0,1\rangle ,|1,0\rangle
\right\} $ are decoupled. In writing Eq.~(\ref{eq:oriZ}), a periodic
boundary condition $n_i^1=n_i^{2M+1}$ is imposed  in the Trotter
direction.

Now let uf introduce the following site-dependent variables:
\begin{eqnarray}
\alpha_i&=&-\frac{\varepsilon(U_i-\mu)}{2}, ~~~~~~~\gamma_i =
\sqrt{\alpha_i^2+\varepsilon^2t_i^2}, \nonumber \\
b_i&=&e^{\alpha_i} , ~~~~~~~~~~~~~~~~~~~ u_i=\frac{\varepsilon
t_i\sinh \gamma_i}{\gamma_i} ,
\nonumber\\
a_i&=&\cosh\gamma_i ,~~~~~~~~~~~w_i=\frac{\alpha_i\sinh
\gamma_i}{\gamma_i}~,\label{eq:para}
\end{eqnarray}
Only three of which are independent. In terms of these variables, we
obtain the following matrix elements for $v_{i,i+1}$ in the state
number representation $\{|00\rangle, |01\rangle, |10\rangle,
|11\rangle\}$:
\begin{equation}
v_{i,i+1}^{l,l+1}=b_i\left( \begin{array}{cccc}
 b_i^{-1}    & 0    & 0    & 0    \\
 0    & a_i-w_i    & u_i   & 0   \\
 0    & u_i    & a_i+w_i    & 0  \\
 0    & 0    & 0    & b_i
 \end{array}
 \right) .
\end{equation}

According to the procedure mentioned above, the resulting local
transfer matrix in the Trotter direction between slices $l$ and
$l+1$ is also block-diagonal because of the fermion number
conservation. Consequently,
\begin{equation}
\tau_{i,i+1}^{l,l+1}=b_i\left( \begin{array}{cccc}
u_i       & 0      & 0       & 0    \\
0       & a_i-w_i    &b_i^{-1}        & 0   \\
0       & b_i  & a_i+w_i  & 0  \\
0       & 0    & 0    & u_i
\end{array}
 \right) . \label{table:T}
\end{equation}
It can be shown that this transfer matrix has the following operator
form:
\begin{eqnarray}
\frac{\tau^{l,l+1}_{i,i+1}}{u_ib_i}&=&1+(\frac{a_i}{u_i}-1)(d^{\dag}_{l}
d_{l}- d^{\dag}_{l+1}
 d_{l+1})^2+\frac{b_i}{u_i}d^{\dag}_{l} d_{l+1}\nonumber\\
 &&+\frac{b_i^{-1}}{u_i}d^{\dag}_{l+1} d_{l}+\frac{w_i}{u_i}(d^{\dag}_{l} d_{l}- d^{\dag}_{l+1}
 d_{l+1}),\label{eq:T}
\end{eqnarray}
where $d$'s are fermion operators defined in the Trotter space.
$\tau^{l,l+1}_{i,i+1}$ is a quadratic function of fermion operators.
Furthermore, $\tau^{l,l+1}_{i,i+1}$ can be exponentiated again to a
concise quadratic form due to the fermion exclusion principle. For
simplicity, we denote $C_i=u_ib_i$,
$s_i\equiv\sqrt{p_iq_i+{r_i}^2}$, $n_l=d^{\dag}_{l} d_{l}$ and
\begin{equation}
A=p_id_l^{\dagger }d_{l+1}+q_id_{l+1}^{\dagger }d_{l}+r_i\left(
n_{l}-n_{l+1}\right),
\end{equation}
then,
\begin{eqnarray}
\frac{\tau_{i,i+1}^{l,l+1}}{C_i} &\equiv& \exp[p_id_l^{\dagger
}d_{l+1}+q_id_{l+1}^{\dagger }d_{l}+r_i\left( n_{l}-n_{l+1}\right)] \nonumber\\
&=&1+(\cosh s_i-1)(n_{l}-n_{l+1})^2+\frac{\sinh s_i}{s_i}A.
\nonumber\\ \label{eq:T-exp}
\end{eqnarray}
This exponential quadratic operator form of $\tau^{i,i+1}_{l,l+1}$
is valid only when the four coefficients before $(n_{l}-n_{l+1})^2$,
$(n_l-n_{l+1})$, $d_l^{\dagger }d_{l+1}$, and $d_{l+1}^{\dagger
}d_{l}$ satisfy the following four equations:
\begin{eqnarray}
\cosh s_i &=&\frac{a_i}{u_i} ,
~~~~~~~\frac{\sinh s_i}{s_i}p_i =\frac{b_i}{u_i} , \nonumber\\
\frac{\sinh s_i}{s_i}q_i &=&\frac{b_i^{-1}}{u_i} ,
~~~~~~~\frac{\sinh s_i}{s_i}r_i =\frac{w_i}{u_i} .
\end{eqnarray}

Now we can cast the column transfer matrix $T_{2i,2i+1}$ defined by
Eq.~(\ref{eq:TaTb}) into the following form:
\begin{eqnarray}
T_{i,i+1}&=&C_{i}^{M}\exp \Big[ \sum_{l=1}^{M}p_id_{2l}^{\dagger
}d_{2l+1}\nonumber\\&&+q_id_{2l+1}^{\dagger }d_{2l}+r_i\left(
n_{2l}-n_{2l+1}\right) \Big] .
\end{eqnarray}

$T_{i,i+1}$ is translational invariant in the Trotter direction.
Therefore, we can introduce a fourier transformation along this
direction:
\begin{equation}
d_{\omega}^{\dagger }=\left(
\begin{array}{c}
d_{\omega,1}^{\dagger } \\
d_{\omega,2}^{\dagger }%
\end{array}%
\right) =\frac{1}{\sqrt{M}}\sum_{l =1}^{M}e^{iwR_{l}}\left(
\begin{array}{c}
d_{2l-1}^{\dagger } \\
d_{2l}^{\dagger }%
\end{array}%
\right) .
\end{equation}
The general column transfer matrix can then be rewritten as

\begin{eqnarray} &T_{2i-1,2i}=C_{2i-1}^{M}\exp
\left[\sum_{\omega}d_{\omega}^{\dagger }\left(
\begin{array}{ll}
r_{2i-1}  & p_{2i-1} \\
q_{2i-1}  & -r_{2i-1}
\end{array}
\right)d_{\omega}\right] , \nonumber\\
&T_{2i,2i+1}=C_{2i}^{M}\exp \left[ \sum_{\omega}d_{\omega}^{\dagger
}\left(
\begin{array}{ll}
-r_{2i}  & e^{-i\omega}q_{2i} \\
e^{i\omega}p_{2i}  & r_{2i}
\end{array}%
\right) d_{\omega}\right] . \nonumber\\
\end{eqnarray}

 Substituting these transfer matrices into Eq.~(\ref{eq:part}),
we obtain the following expression for the partition function:
\begin{equation}
Z=\prod_{i}C_i^M\prod_{\omega} {\rm Tr} \left[2+ T_\omega \right] ,
\label{eq:Z1}
\end{equation}
where
\begin{equation}
T_\omega =\prod_{i=1}^{N/2}t_{2i-1} t_{2i,\omega}, \label{eq:T_q}
 \end{equation}
$t_{2i-1}$ and $t_{2i,\omega}$ are $2\times 2$ matrices defined by
 \begin{eqnarray}
t_{2i-1}&=&\frac{1}{u_{2i-1}}\left(\begin{array}{ll}
a_{2i-1}-w_{2i-1}  & b_{2i-1}^{-1} \\
b_{2i-1} & a_{2i-1}+w_{2i-1}
\end{array}
\right) , \nonumber \\
t_{2i,\omega}&=&\frac{1}{u_{2i}}\left(
\begin{array}{ll}
a_{2i}+w_{2i}  & e^{-i\omega}b_{2i} \\
e^{i\omega}b_{2i}^{-1}  &a_{2i}-w_{2i}
\end{array}
\right) .
\end{eqnarray}
In obtaining these expressions, we have used the fact that
$\tau^{i,i+1}_{l,l+1}/C_i$ is an identity matrix in the subspace $
\left\{ |1,1\rangle \right\} $, $\left\{ |0,0\rangle \right\} $. In
Eq.~(\ref{eq:T_q}), all multiplied matrices are exponential of
traceless matrices. Thus the eigenvalue of the final matrix after
multiplications will have the form $\exp[\pm\lambda(\omega)]$. In
the thermodynamic limit, the constant $2$ in Eq.~(\ref{eq:T_q}) can
be neglected because the eigenvalue $\exp[\lambda(\omega)]$
dominates.

It should be noted that $\omega$ is related to the parity of $M$.
For odd $M$
\begin{equation}
 \omega=\frac{2m\pi}{M},\qquad
 m=-\frac{M-1}{2},...0,...\frac{M-1}{2}, \label{eq:oddq}
\end{equation}
and for even $M$
\begin{equation}
\omega=\frac{(2m+1)\pi}{M},\qquad
 m=-\frac{M}{2},...-1,0,...\frac{M}{2}-1. \label{eq:evenq}
\end{equation}
The transfer matrix $T_\omega$ in Eq.~(\ref{eq:T_q}) is calculated
in the subspace of $ \left\{ |0,1\rangle , |1,0\rangle \right\}$, so
the fermion occupation number is 1 for each two unit cells along the
Trotter direction. The state space for one column transfer matrix is
\begin{equation}
d_1^{\dagger}d_2^{\dagger}...d_{2M-1}^{\dagger}
d_{2M}^{\dagger}|0\rangle £¬ \end{equation} and the total occupation
number is $M$. Due to the periodicity in Trotter direction, it is
equivalent to calculate in the state space
\begin{equation}d_2^{\dagger}d_3^{\dagger}...d_{2M}^{\dagger}
d_{2M+1}^{\dagger}|0\rangle.
\end{equation} However, permuting the fermion operator $d_1^{\dagger}$ behind
$d_{2M}^{\dagger}$ will bring factor $(-1)^{M-1}$ because of the
transposition with $(M-1)$ occupied fermions. Therefore, there are
the relation,
\begin{equation}
(-1)^{M-1}d_1^{\dagger}=d_{2M+1}^{\dagger}.
\end{equation} For odd $M$, $d_1^{\dagger}=d_{2M+1}^{\dagger}$; For even $M$,
$d_1^{\dagger}=-d_{2M+1}^{\dagger}$. As a result, $\omega$ takes
values according to the formulas (\ref{eq:oddq}) and
(\ref{eq:evenq}) respectively. In the following, we will set $M$ to
even and use Eq.~(\ref{eq:evenq}).

As shown in Ref. \cite{Shankar}, $T_{\omega}$ in Eq.~(\ref{eq:T_q})
can be formally regarded as the transfer matrix for the following 1D
Ising model in a magnetic field:
\begin{equation}
Z=\sum_{S_i}\exp\{\sum_i\left[J_i(S_iS_{i+1}-1)+h_iS_i+f_i\right]\}
.
\end{equation}
For the odd sites, the parameters are $\omega$-independent in the
corresponding 1D Ising model with the relations
\begin{eqnarray}
f_{2i-1}&=&-\frac{1}{2}(\ln|b_{2i-1}|) , \\
h_{2i-1}&=&\ln\frac{(a_{2i-1}-w_{2i-1})}{(b_{2i-1}+w_{2i-1})} , \\
J_{2i-1}&=&\frac{1}{4}\ln\frac{2u_{2i-1}}{|b_{2i-1}|} .
\end{eqnarray}

However, the situation is different for even sites and the
parameters $J_i$, $f_i$ are now $\omega$-dependent
\begin{eqnarray}
f_{2i}&=&\frac{1}{2}(\ln|b_{2i}|+i\omega) , \\
h_{2i}&=&\ln\frac{(a_{2i}+w_{2i})}{(a_{2i}-w_{2i})} , \\
J_{2i}&=&\frac{1}{4}(2\ln (u_{2i}|b_{2i}|)+i\omega) .
\end{eqnarray}

The correspondence for even sites cannot provide a material mapping
to 1D Ising model because of the complex parameters, which arises
from the fourier transformation along the Trotter direction. Since
there exists a relation $T^*_{\omega}=T_{-\omega}$ due to the factor
$e^{i\omega}$, we can multiply in pair the transfer matrices with
$\omega$ and $-\omega$. In fact, the pairwise multiplication can
save us half of the calculation because of the complex conjugate
relation. When using Eq.~(\ref{eq:T_q}) to solve the partition
function, what is needed is to calculate the product of $N$ matrices
of dimension $2$ for a given $\omega$. When $Z$ is obtained, the
calculation of the free energy is direct, $F=-T\ln Z$ (hereafter
taking $k_B=1$), with which all other thermodynamic quantities
interested such as the specific heat $C$ can be obtained.

\begin{figure}[ht]
\includegraphics [width=0.5\textwidth] {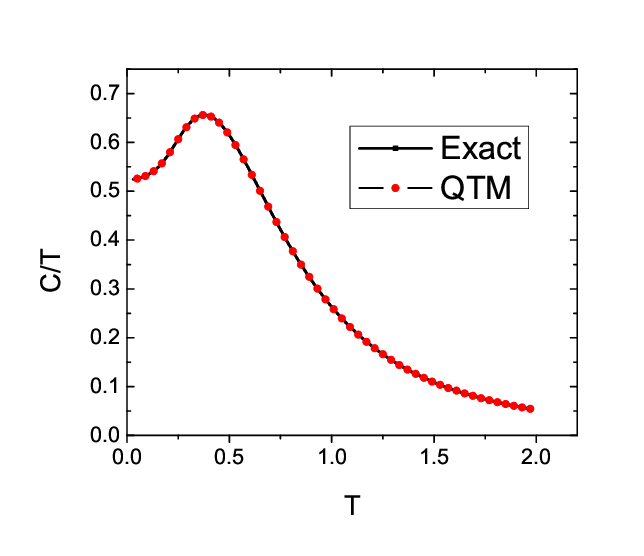}
\caption{(Color online) Comparison of the specific heat coefficient
$C/T$ obtained from two different methods: the exact energy spectrum
calculation and the quantum transfer matrix method. Here, $t=1$ and
$U=0$.} \label{fig:C/T}
\end{figure}

Fig.~\ref{fig:C/T} shows the result of this comparison for the
linear coefficient of the specific heat $C/T$. It is obvious that
the two sets of data coincide completely. The chain length $N$ here
and in the rest of the paper is taken up to $2^{17}(=131072)$.

\section{Results}
\subsection{Gaussian diagonal disorder}

We first consider a Gaussian-like function
\begin{equation}
P(x)=\frac{1}{\sqrt{2\pi}\sigma}e^{\frac{-(x-a)^2}{2\sigma^2}}
\end{equation}
for the disordered distribution of the diagonal potential $U_i$,
where $a$ and $\sigma$ are the mean value and the standard deviation
respectively. In the following discussion, we take $t=1$, $a=0$,
$\mu=0$. The controlling  parameter is the standard deviation
$\sigma$, which denotes the disorder degree of the distribution of
$U_i$. In these figures, the red solid line means the case without
disorder, i.e. $U_i=0$ ($\sigma=0$), and the other curves are for
$\langle U_i \rangle=a=0$ and $\sigma\neq0$.
\begin{figure}
\includegraphics[width=0.5\textwidth]
{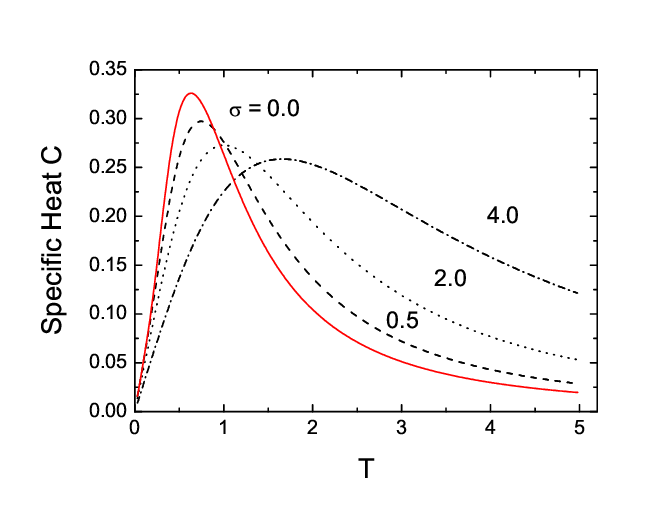} \caption {(Color online) The specific heat $C$ as a
function of $T$ for $\sigma= 0.5$, 2, 4, and $0$(red solid line).}
\label{fig:Mu0diffW-C}
\end{figure}

\begin{figure}
\includegraphics [width=0.5\textwidth]
{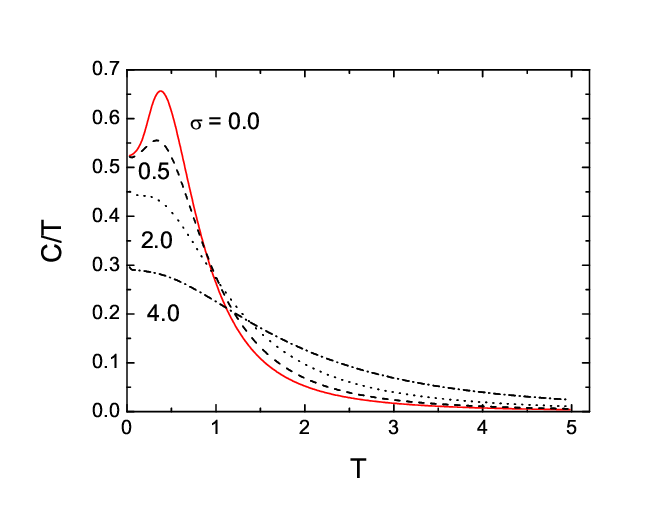} \caption{(Color online) The specific heat
coefficient $C/T$ as a function of $T$ for $\sigma= 0.5$, 2, 4, and
$0$(red solid line).} \label{fig:Mu0diffW-CT}
\end{figure}

\begin{figure}
\includegraphics [width=0.5\textwidth] {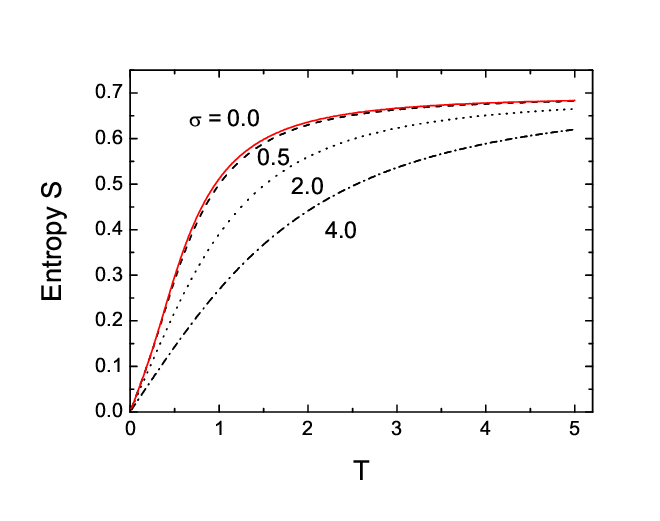}
\caption{(Color online) The entropy $S$ as a function of $T$ for
$\sigma= 0.5$, 2, 4, and $0$(red solid line).}
\label{fig:Mu0diffW-S}
\end{figure}

As shown in Figs.~\ref{fig:Mu0diffW-C}-\ref{fig:Mu0diffW-S}, the
difference of the physical quantities is slight between the cases of
$\sigma = 0.5$ and $\sigma = 0$. However, significant differences
appear when $\sigma$ increases.  With increasing $\sigma$, the peaks
of the specific heat $C$ move towards higher temperatures. The
introduction of disordered diagonal energy widens the energy band.
The density of states(DOS) distributes in a broaden energy range.
This leads to the shift of the peak position of $C$ towards higher
temperature with increasing $\sigma$. The disorder assists the
thermal fluctuations and shifts the peak of $C/T$ to lower
temperatures. The specific heat coefficient $C/T$ at zero
temperature is proportional to the density of states around the
Fermi energy $E_F$, i.e.,
\begin{equation}
\frac{C}{T}\mid_{T\rightarrow0}~\varpropto\rho(E_F).
\end{equation}
Fig.~\ref{fig:Mu0diffW-CT} shows that when the disorder increases,
the density of state near the fermi surface decreases. The entropy
$S$ decreases when $\sigma$ increases.

\begin{figure}
\includegraphics [width=0.5\textwidth]{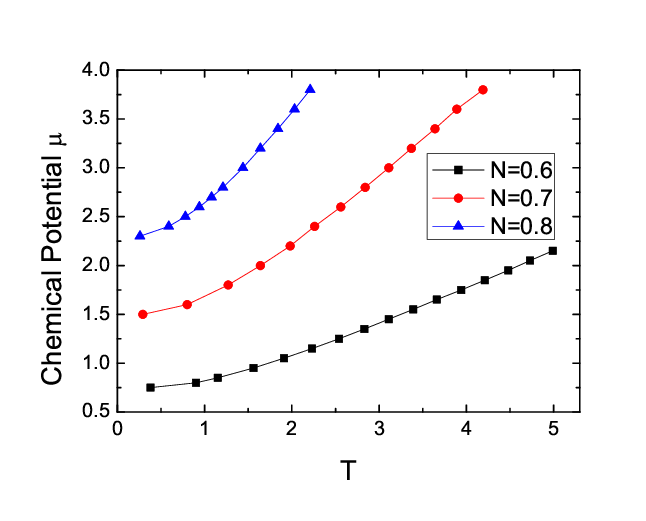}
\caption {(Color online) The chemical potential $\mu$ as a function
of $T$ for some fixed occupation numbers $N=0.6,0.7,0.8$. $a=0$,
$\sigma=2$, and $t=1$. } \label{fig:Mu-T}
\end{figure}

Fig.~\ref{fig:Mu-T} shows how the chemical potential $\mu$ changes
with $T$ for some given occupation numbers $N$. $N$ becomes large
with $\mu$ increasing when $T$ keeps invariant, while $N$ become
small with $T$ increasing when $\mu$ keeps intact. To keep $N$
invariant, $\mu$ is a monotonic increasing function of $T$. The
function shows linear shape at high temperature regime.

\begin{figure}
\includegraphics [width=0.5\textwidth] {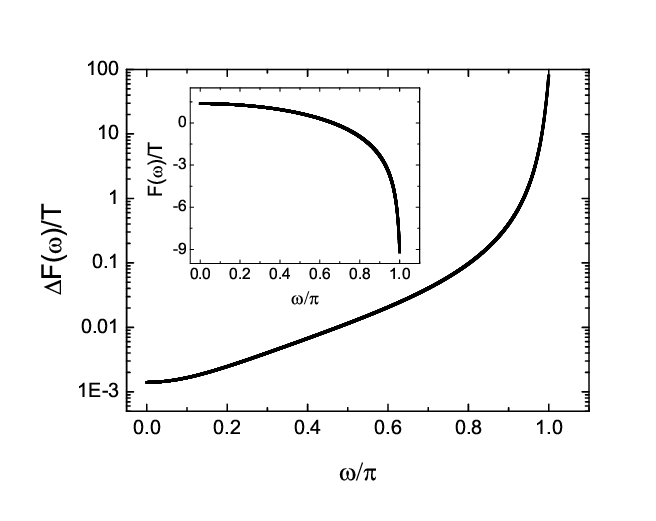}
\caption{The difference $\Delta F(\omega)/T$ between disordered and
uniform cases as a function of $\omega/\pi$. Here, $T=0.01$,
$t=U=1,~\sigma=2$. The inset shows the $\omega$ dependence of
$F(\omega)/T$ for the system without disorder.} \label{fig:delta-fq}
\end{figure}

In Eq.~(\ref{eq:Z1}), the partition function is expressed as product
of different $\omega$ components, consequently, the free energy $F$
can be written as a sum of dependent free energy $F(\omega)$. For a
given temperature $T$, $F(\omega)$ decreases with $\omega$
increasing. When the disorder is turned on, $F(\omega)$ changes
little except in the vicinity of $\omega \sim \pi$. This can be seen
from Fig.~\ref{fig:delta-fq}, which compares the difference of free
energies between the disordered ($\sigma=2$) and ordered cases as a
function of $\omega$. It clearly shows that the difference becomes
significant only when $\omega$ approaches $\pi$. In Ref.
\cite{Shankar}, the singularity from some special $\omega$'s was
used to discuss the phase transition.

\subsection{Staggered disorder potential}

We now consider a special model whose diagonal potential energy is
alternating (staggered) with the lattice site, i.e.,
\begin{equation} t_i=t~,~~~U_i = (-1)^i U + \Delta U_i.
\end{equation} When $\Delta U_i =0$, the energy spectrum is readily
calculated via the Fourier transformation,
\begin{equation}
c_i=\sum_k e^{iki}c_k, ~~~~~~c_i^\dagger=\sum_k e^{-iki}c_k^\dagger.
\end{equation}
The result is two sub-bands dispersion relation,
\begin{equation}
E_{\pm}=-2\mu\pm\sqrt{U^2+4t^2\cos^2k}~.
\end{equation}
The band gap is $2U$.

Let us choose a uniform rectangular distribution
$P(x)=\frac{1}{\sigma}$ for the increment of the random diagonal
energy $\Delta U_i$. Here, $\sigma$ is the width of the rectangular
distribution satisfying
\begin{equation}
P(x)=\left\{\begin{array}{cl} \frac{1}{\sigma},~~&~~{\rm
if}~~~~-\frac{\sigma}{2}\leq x\leq\frac{\sigma}{2},\vspace{0.4cm}\\
 0,~~&~~{\rm otherwise.}
\end{array}
\right.
\end{equation}

\begin{figure}
\includegraphics [width=0.5\textwidth]
{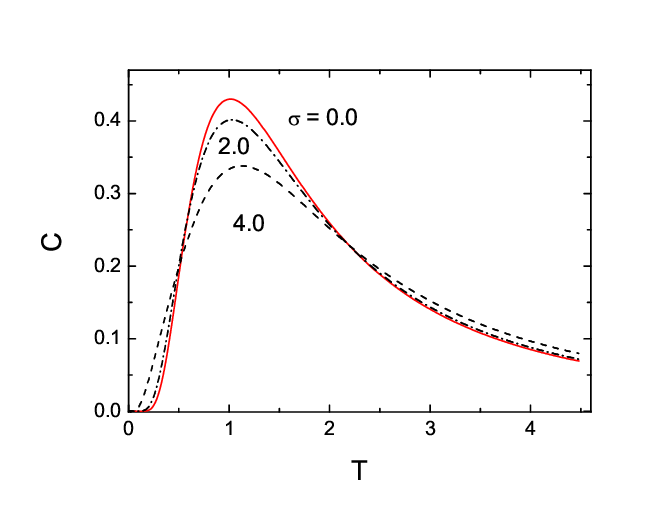} \caption {(Color online) The specific heat $C$ as a
function of $T$ for the uniform($\sigma=0$) and
disordered($\sigma=2,4$) potentials. $t=1,U=2,\mu=0$}
\label{fig:two-C}
\end{figure}

In the presence of disorder, the band gap is expected to decrease
with the increase of the disorder degree $\sigma$. For small
$\sigma$, there is a finite excitation gap and the specific heat
drops exponentially in low temperature, as shown in
Fig.~\ref{fig:two-C} .

We also consider the case when the random potential take only two
discrete values: $-\sigma/2,\sigma/2$ with equal probability. We
call it discrete distribution of the increment, and the above
rectangular distribution is denoted by continuous one.

\begin{figure}
\includegraphics [width=0.5\textwidth] {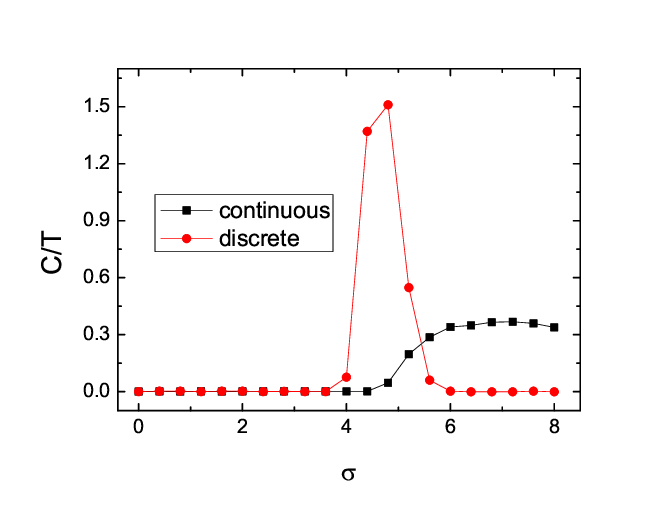}
\caption{(Color online) Comparison of the specific heat coefficient
$C/T$ for a continuous distributed random potential with that for a
discrete random potential. $t=1,U=2,\mu=0$ and $T=0.01$. }
\label{fig:dis-con}
\end{figure}

Fig.~\ref{fig:dis-con} compares the specific heat coefficient $C/T$
at $T=0.01$ for the above two kinds of distribution of random
potentials. In the case of discrete distribution, $C/T$ exhibits a
sharper peak. When the disordered level $\sigma$ increases to
approximately $4$, the band gap disappears. With further increasing
$\sigma$, $C$ drops to zero because the band gap open again for the
discrete random potential. This can be understood as follows. The
two sub-bands close to each other when disorder is introduced. The
upper sub-band shifts downwards by $\sigma/2$ and the lower sub-band
shifts upwards by $\sigma/2$. When the top of the original upper
sub-band touches the zero energy, i.e., $\sigma/2=\sqrt{6}$, the two
sub-bands begin to separate again. Therefore, $C/T$ decreases to
zero at $\sigma\simeq 2\sqrt{6}$ in Fig.~\ref{fig:dis-con}. On the
contrary, for the continuous random increment case, $C/T$ remains a
finite value even for large $\sigma$. This is because the splitting
of the upper and lower bands only blurringly expands the width of
these two bands, and once they touch each other, they never separate
again.

As a further investigation for $C$ in discrete distribution, we
choose three typical disordered degrees $\sigma=3.0,4.8,7.0$,
corresponding to three regions in Fig.~\ref{fig:dis-con}, to show
how $C$ varies with the temperature at low $T$.
\begin{figure}
\includegraphics [width=0.5\textwidth] {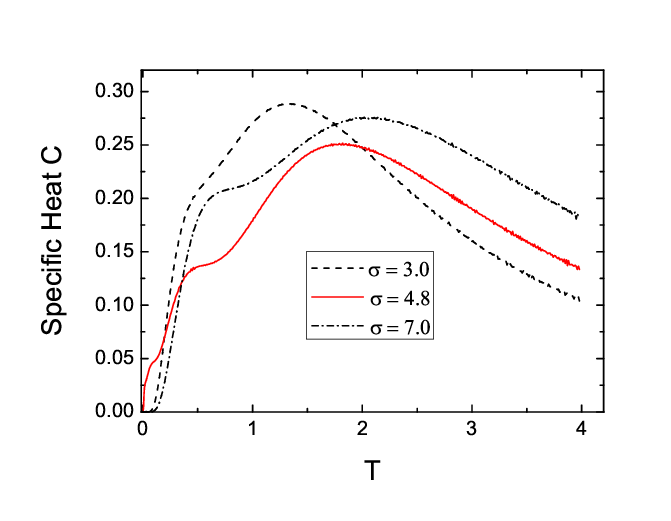}
\caption{(Color online) Temperature dependence of the specific heat
in three different random potentials($\sigma=3.0,4.8,7.0$).
$t=1,U=2,\mu=0$.} \label{fig:two-Ctwo}
\end{figure}
As shown in Fig.~\ref{fig:two-Ctwo}, when $\sigma=3.0$ and 7.0, $C$
drops exponentially with temperature in low temperatures. The
double-peak structure of $C$ can be understood from the overlaps of
energy bands. The discrete increments split the original single
energy band into the two bands which shift upwards and downwards by
$\sigma/2$. The resulting four bands from upper and lower sub-bands
meet pairwise. The thermal fluctuations bring double peaks shown in
$C$ curves.

\section{conclusion}

For the 1D disordered system, the quantum transfer matrix method we
have developed is applicable to all kinds of disorder distribution
types and strengths. The non-diagonal disordered problem can be
handled since the partition function $Z$ can be expressed as the
product of site-dependent local transfer matrices. Compared to the
diagonal disordered cases, we only need to modify the local transfer
matrix elements correspondingly.

We have studied the thermodynamic properties of the 1D disordered
Anderson model. We discussed two kinds of diagonal (potential)
disordered models with or without staggered potentials. The free
energy $F$ can be written as a sum of different $\omega$ components
from the Fourier transformation in Trotter space. Comparing with the
system without disorder, the most significant difference in
$F(\omega)$ shows only in the region very close to $\omega=\pi$. The
disorder change the distribution of DOS, leading to the difference
in the thermodynamic quantities in comparison with the disorder-free
system. All the results shown in the figures are for systems with
the number of sites greater than $10^5$. This kind of calculation is
far beyond the capacity of exact diagonalization.

The transfer matrix method has a broad range of applicability and
can be used to discuss any non-interacting fermion models. Recently
this method has been used to calculate the thermodynamic quantities
of the Hofstadter model which describes the behavior of
tightly-bound Bloch electrons in the magnetic field
\cite{Wenhu,liping}. In Landau gauge, the Hofstadter Hamiltonian can
be decoupled into a sum of one dimensional Hamiltonian, which lies
in the application range of the quantum transfer matrix method.

\acknowledgments

This work was supported by the National Natural Science Foundation
of China and the National Program for Basic Research of MOST, China.


\begin{thebibliography}{99}
\bibitem{Lee}P. A. Lee and T. V. Ramakrishnan, Rev. Mod. Phys. {\bf 57},
287 (1985).

\bibitem{Anderson58}P. W. Anderson, Phys. Rev. {\bf 109}, 1492 (1958).

\bibitem{Altshuler}B. L. Altshuler and A. G. Aronov, Solid State
Commun. {\bf 39}, 115 (1979).

\bibitem{Wegner}F. Wegner, Z. Phys. B {\bf 25}, 327 (1976).

\bibitem{Abrahams}E. Abrahams, P. W. Anderson, D. C. Licciardello, and T. V. Ramakrishnan,
Phys. Rev. Lett. {\bf 42}, 673 (1979).

\bibitem{Brandes}T. Brandes and S. Kettemann, \emph{Anderson Localization
and Its Ramifications}, (Springer, Berlin, 2003).

\bibitem{Tao}R. J. Bursill, T. Xiang, and G. A. Gehring, J.Phys.:
Condens. Matter {\bf 8}, L583 (1996).

\bibitem{Tao1}X. Q. Wang and T. Xiang, Phys. Rev. B {\bf 56}, 5061 (1997).

\bibitem{Xiang}T. Xiang and X. Wang, in \emph{Density-Matrix Renormalization: A
New Numerical Method in Physics}, edited by I. Peschel, X. Wang, M.
Kaulke, and K. Hallberg (Springer, New York, 1999), pp. 149-172.

\bibitem{Trotter}H. F. Trotter, Proc. Am. Math. Soc. {\bf 10},
545 (1959).

\bibitem{Suzuki}M. Suzuki,
Prog. Theor. Phys. {\bf 56}, 1454 (1976).

\bibitem{Betsuyaku}H. Betsuyaku,
Prog. Theor. Phys. {\bf 73}, 320 (1985).

\bibitem{Shankar}R. Shankar and G. Murthy, Phys. Rev. B {\bf
36}, 536 (1987).

\bibitem{Wenhu}W. H. Xu, L. P. Yang, M. P. Qin and T. Xiang,  arXiv: 0808.0099 (2008).

\bibitem{liping}L. P. Yang and T. Xiang, unpublished.

\end{thebibliography}
\end{document}